\documentclass[a4paper,11pt]{article}

\usepackage{contribution}

\newcommand{\weblink}[2][]{%
    \ifthenelse{\equal{#1}{}}%
    {\textnormal{\url{#2}}}%
    {\textnormal{\href{#2}{#1}}}%
}

\newcommand{\acknowledgements}[1]{%
  \bigskip\bigskip
  \textsf{\textbf{\Large Acknowledgements}} \\[2ex]
  {#1}
  \bigskip
}

\def\beq{\begin{equation}}
\def\eeq#1{\label{#1}\end{equation}}
\def\eeqn{\end{equation}}

\def\beqa{\begin{eqnarray}}
\def\eeqa#1{\label{#1}\end{eqnarray}}
\def\eeqan{\end{eqnarray}}

\let\bar=\overbar

\def\Dslash{\not{\hbox{\kern-4pt $D$}}}
\def\dslash{\not{\hbox{\kern-2pt $\del$}}}

\def\msb{{\bar{\ssstyle M \kern -1pt S}}}

\newcommand{\contribution}[7][]{%
  \clearpage
  \thispagestyle{plain}
  \ifthenelse{\equal{#1}{}}
  {\hypersetup{pdftitle={#2}}}
  {\hypersetup{pdftitle={#1}}}
  \hypersetup{pdfauthor={{#3} {#4}}}
  {\centering\normalfont\LARGE\bfseries\sffamily #2 \par\nobreak}
  \lhead{}
  \chead{%
    \textit{\footnotesize XIV International Conference on Hadron Spectroscopy
      (\weblink[\textit{hadron2011}]{http://www.hadron2011.de}), 13-17 June 2011, Munich, Germany}%
  }
  \rhead{}
  \bigskip
  \begin{center}
    {#3} {#4}\ifthenelse{\equal{#6}{}}{}{\footnote{\weblink[#6]{mailto:#6}}}
    \ifthenelse{\equal{#7}{}}{}{#7} \\
    \textit{#5}
  \end{center}
  \bigskip
}

\renewcommand{\abstract}[1]{%
  \begin{center}
    \begin{minipage}{0.85\textwidth}
      \begin{footnotesize}
        #1
      \end{footnotesize}
    \end{minipage}
  \end{center}
  \bigskip
}

\begin{document}

%
%
%
%
%
{  

\makeatletter
\@ifundefined{c@affiliation}%
{\newcounter{affiliation}}{}%
\makeatother
\newcommand{\affiliation}[2][]{\setcounter{affiliation}{#2}%
  \ensuremath{{^{\alph{affiliation}}}\text{#1}}}

\renewcommand{\beq}{\begin{equation}}
\renewcommand{\eeq}{\end{equation}}
\newcommand{\Fpi}{F_\pi}
\newcommand{\mpi}{M_{\pi}}
\newcommand{\ga}{g_{\rm A}}
\newcommand{\diff}{\text{d}}
\newcommand{\Order}{\mathcal{O}}
\newcommand{\tm}{t_{\rm m}}
\newcommand{\sm}{s_{\rm m}}
\newcommand{\qq}{\mathbf{q}}
\newcommand{\qt}{\mathbf{q}_t}
\newcommand{\pt}{\mathbf{p}_t}
\newcommand{\Lagr}{\mathcal{L}}
\renewcommand{\Im}{\text{Im}\,}
\renewcommand{\Re}{\text{Re}\,}
%

\contribution[Roy--Steiner equations for $\gamma\gamma\to\pi\pi$]  
{Roy--Steiner equations for $\boldsymbol{\gamma\gamma\to\pi\pi}$}  
{Martin}{Hoferichter}  
{\affiliation[Helmholtz-Institut f\"ur Strahlen- und Kernphysik (Theorie) and]{1}\\
 Bethe Center for Theoretical Physics, Universit\"at Bonn, Germany \\
\affiliation[Institute of Nuclear and Particle Physics and Department of Physics and Astronomy,]{2}\\
Ohio University, Athens, USA\\
\affiliation[Instituto de F\'{\i}sica de Buenos Aires, CONICET - Departamento de F\'{\i}sica, FCEyN,]{3}\\
Universidad de Buenos Aires, Argentina}  
{hoferichter@hiskp.uni-bonn.de}  
{\!\!$^,\affiliation{1}^,\affiliation{2}$, Daniel R.~Phillips\affiliation{2}, and Carlos Schat$\affiliation{2}^,\affiliation{3}$}

%

\abstract{%
  Starting from hyperbolic dispersion relations, we present a system of Roy--Steiner equations for pion Compton scattering that respects analyticity and unitarity requirements, gauge invariance, as well as crossing symmetry, and thus all symmetries of the underlying quantum field theory. To suppress the dependence on the high-energy region, we also consider once- and twice-subtracted versions of the equations, where the subtraction constants are identified with dipole and quadrupole pion polarizabilities. We consider the resolution of the $\gamma\gamma\to\pi\pi$ partial waves by a Muskhelishvili-Omn\`es representation with finite matching point, and discuss the consequences for the two-photon coupling of the $\sigma$ resonance as well as its relation to pion polarizabilities.
}
%

\section{Introduction}

The Roy equations for $\pi\pi$ scattering \cite{Roy} are a coupled system of partial wave dispersion relations that respects analyticity, unitarity, and crossing symmetry of the scattering amplitude. In recent years, partial wave dispersion relations in combination with unitarity (and chiral symmetry) have been used for high-precision studies of low-energy processes, both in $\pi\pi$ \cite{ACGL,madrid} and $\pi K$ \cite{piK} scattering. An important application of $\pi\pi$ Roy equations in combination with Chiral Perturbation Theory (ChPT) was the precise prediction of the pole parameters of the $\sigma$ resonance \cite{CCL}
\beq
M_\sigma=441^{+16}_{-8}\,{\rm MeV},\qquad \Gamma_\sigma=544^{+18}_{-25}\,{\rm MeV}.
\eeq
The reaction $\gamma\gamma\to\pi\pi$ provides an alternative to $\pi\pi$ scattering for the excitation of the $\sigma$. In particular, as discussed in detail in \cite{HPS_11}, Roy-equation techniques in $\gamma\gamma\to\pi\pi$ allow us to constrain the $\sigma$'s two-photon width $\Gamma_{\sigma\gamma\gamma}$ at a similar level of rigor as $M_\sigma$ and $\Gamma_\sigma$ based on $\pi\pi$ Roy equations.

\section{Roy equations for $\boldsymbol{\pi\pi}$ scattering}

Roy equations for $\pi\pi$ scattering are obtained by starting from a twice-subtracted dispersion relation at fixed Mandelstam $t$, determining the $t$-dependent subtraction constants by means of crossing symmetry, and finally performing a partial wave expansion. This leads to a coupled system of integral equations for the $\pi\pi$ partial waves $t^I_J(s)$ with isospin $I$ and angular momentum $J$
\beq
\label{pipi_roy}
t_J^I(s)=k_J^I(s)+\sum\limits_{I'=0}^{2}\sum\limits_{J'=0}^\infty\int\limits_{4\mpi^2}^\infty \text{d}s'K_{JJ'}^{II'}(s,s')\Im t_{J'}^{I'}(s')
\eeq
where $K_{JJ'}^{II'}$ are known kinematical kernel functions and the $\pi\pi$ scattering lengths---the only free parameters---appear in the subtraction term $k_J^I$. Assuming elastic unitarity
\beq
\Im t_{J}^{I}(s)=\sigma(s)|t_{J}^{I}(s)|^2,\quad t_{J}^{I}(s)=\frac{e^{2i\delta^I_J(s)}-1}{2i\sigma(s)}, \quad \sigma(s)=\sqrt{1-\frac{4\mpi^2}{s}},
\eeq
\eqref{pipi_roy} translates into a coupled integral equation for the phase shifts $\delta^I_J$ themselves.

\section{Roy--Steiner equations for $\boldsymbol{\gamma\gamma\to\pi\pi}$}

Crossing symmetry in this case is less restrictive than for $\pi\pi$ scattering, as it couples $\gamma\gamma\to\pi\pi$ to pion Compton scattering $\gamma\pi\to\gamma\pi$, which we will consider as the $s$-channel process. Roy--Steiner equations are then most conveniently constructed based on hyperbolic dispersion relations \cite{HS_73}. The resulting system of integral equations couples the $\gamma\gamma\to\pi\pi$ partial waves  $h^I_{J,\pm}(t)$ to the $\gamma\pi\to\gamma\pi$ partial waves $f^I_{J,\pm}(s)$ (with photon helicities $\pm$), e.g.
\beq
h^I_{J,-}(t)=\tilde N_J^-(t)+\frac{1}{\pi}\int\limits_{\mpi^2}^\infty\diff s'\sum\limits_{J'=1}^\infty \tilde G_{JJ'}^{-+}(t,s')\Im f^I_{J',+}(s')+\frac{1}{\pi}\int\limits_{4\mpi^2}^\infty\diff t' \sum_{J'}\tilde K_{JJ'}^{--}(t,t')\Im h^I_{J',-}(t'),
\eeq
where $\tilde N_J^-(t)$ includes the QED Born terms. Subtracting at $t=0$, $s=\mpi^2$, the subtraction constants directly correspond to pion polarizabilities. In the once-subtracted case, one needs the dipole polarizabilities $\alpha_1\pm\beta_1$, while a second subtraction requires in addition knowledge of the quadrupole polarizabilities $\alpha_2\pm\beta_2$.

Elastic unitarity is also less restrictive than for $\pi\pi$ scattering, since the unitarity relation is linear in $h^I_{J,\pm}$
\beq
\Im h^I_{J,\pm}(t)=\sigma(t)h^I_{J,\pm}(t)t_{J}^{I}(t)^*.
\eeq
Below inelastic thresholds the phase of $h^I_{J,\pm}$ coincides with $\delta^I_J$ (``Watson's theorem''). Assuming this phase to be known, the equations thus reduce to a Muskhelishvili--Omn\`es problem for $h^I_{J,\pm}$ \cite{MuskhelishviliOmnes}.

\section{Muskhelishvili--Omn\`es solution and results for $\boldsymbol{\Gamma_{\sigma\gamma\gamma}}$}

To solve the equations for $h^I_{J,\pm}$, we truncate the system at $J=2$. Furthermore, we assume the amplitudes to be known above the matching point $\tm=(0.98\,{\rm GeV})^2$. The solution can then be written down in terms of Omn\`es functions
\beq
\Omega^I_J(t)=\exp\Bigg\{\frac{t}{\pi}\int\limits_{4\mpi^2}^{\tm}\diff t'\frac{\delta^I_J(t')}{t'(t'-t)}\Bigg\}.
\eeq
We find that the solutions for different partial waves in general do not decouple, e.g.~the equation for the $S$-wave involves spectral integrals over the $D$-waves as well \cite{HPS_11}. This is a new result of our dispersive treatment of $\gamma\gamma\to\pi\pi$ based on Roy--Steiner equations.

We approximate $\Im f^I_{J,\pm}(s)$, which at low energies is dominated by multi-pion states, by a sum of resonances \cite{GM_10}. Above the matching point we use a Breit--Wigner description of the $f_2(1270)$, which dominates the cross section at higher energies. 
Within our formalism~\cite{HPS_11} we derive a sum rule for the $I=2$ polarizabilities, which---in combination with ChPT results for dipole and neutral-pion quadrupole polarizabilities~\cite{GIS}--- produces an improved prediction 
\beq
(\alpha_2-\beta_2)^{\pi^\pm}=(15.3\pm 3.7)\cdot 10^{-4}\text{fm}^5
\eeq
for the charged-pion quadrupole polarizability. This sum-rule result together with the ChPT values for the other polarizabilities~\cite{GIS} leads to the ``ChPT'' prediction for the total cross section of $\gamma\gamma\to\pi^0\pi^0$ depicted in the left panel of Fig.~\ref{fig:results}. The result labeled ``GMM'' is found when we adopt the polarizability values of a recent fit of a two-channel Muskhelishvili--Omn\`es representation to $\gamma\gamma\to\pi\pi$ cross section data \cite{GM_10}. 
The uncertainty due to the $\pi\pi$ phases represented by the grey band is estimated by varying between two recent state-of-the-art analyses based on Roy and Roy-like equations~\cite{madrid,bern}.
We see that especially for the twice-subtracted version the agreement with experiment in the low-energy region is very good. 
Since we have shown that the $\sigma$ lies within the domain of validity of our Roy--Steiner equations~\cite{HPS_11}, this formalism allows for a reliable analytic continuation to the $\sigma$ pole.

The main result of our analysis is shown in the right panel of Fig.~\ref{fig:results}: there is a correlation between $\Gamma_{\sigma\gamma\gamma}$ and the $I=0$ pion polarizabilities that follows from Roy--Steiner equations and input for the $\pi\pi$ phases alone. In combination with the ChPT-plus-sum-rule input for 
the polarizabilities, we obtain
\beq
\Gamma_{\sigma\gamma\gamma}=(1.7\pm 0.4)\,\text{keV}.
\eeq

\begin{figure}[htb]
  \begin{center}
   \includegraphics[width=0.495\textwidth]{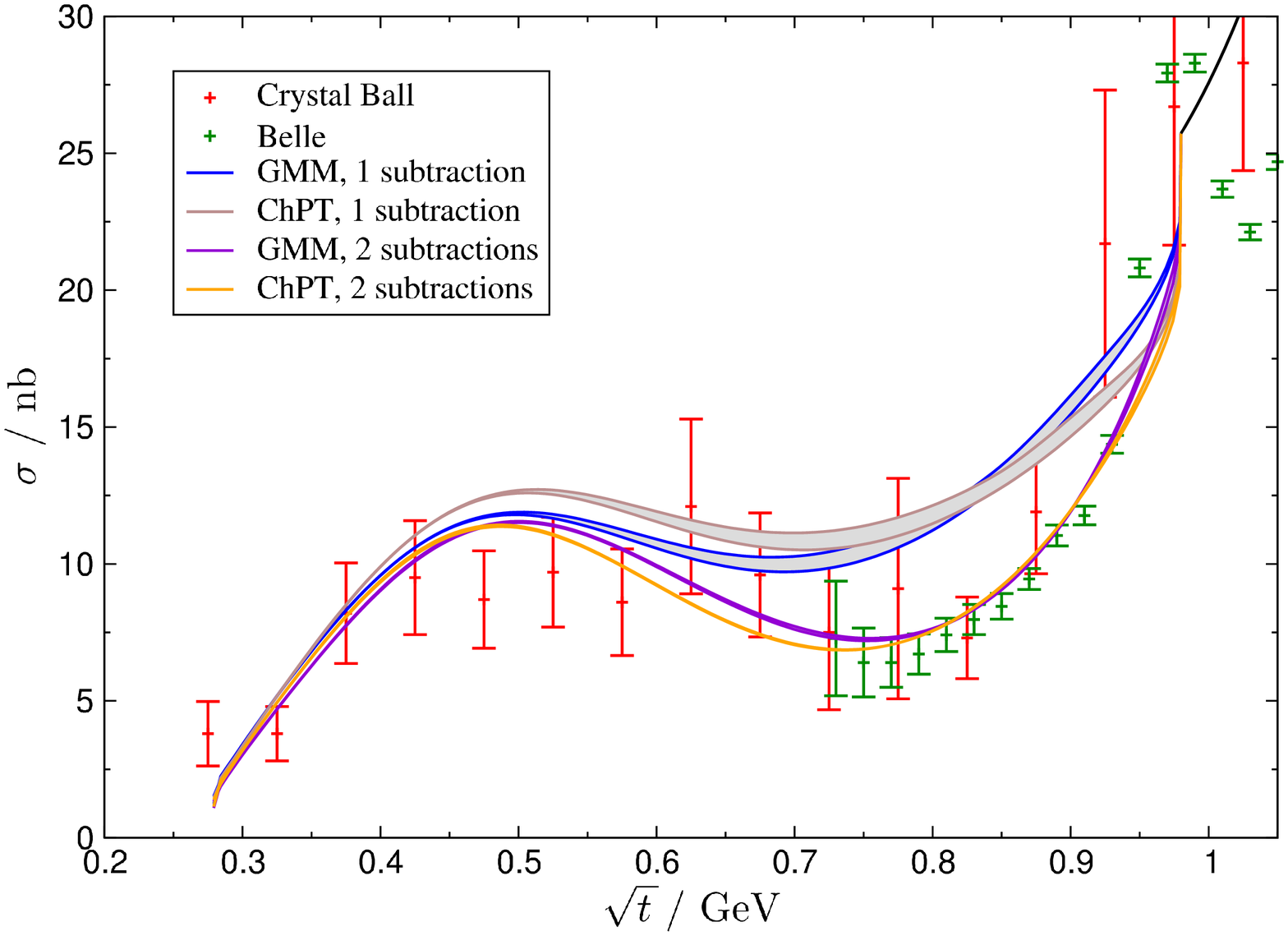}
    \includegraphics[width=0.495\textwidth]{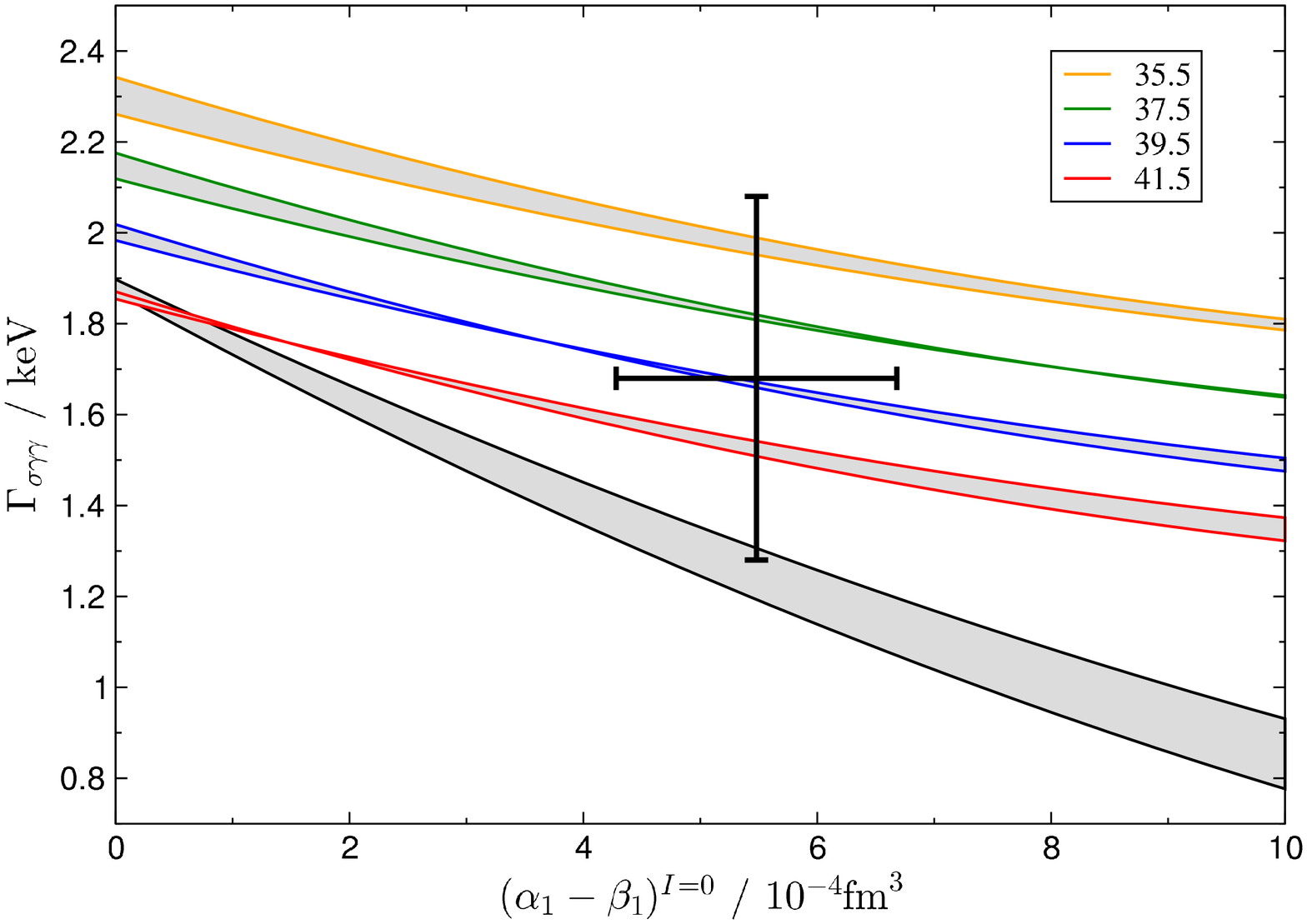}
   \caption{Total cross section for $\gamma\gamma\to\pi^0\pi^0$ for $|\cos\theta|\leq 0.8|$ (left) and $\Gamma_{\sigma\gamma\gamma}$ as a function of the $I=0$ pion polarizabilities (right). The black line refers to the unsubtracted case and the colored lines to the twice-subtracted version with $(\alpha_2-\beta_2)^{I=0}$ as indicated (in units of $10^{-4}{\rm fm}^5$). The grey bands represent the uncertainty due to the $\pi\pi$ input. The cross corresponds to the twice-subtracted case plus ChPT input.}
   \label{fig:results}
  \end{center}
\end{figure}

\newpage

\acknowledgements{%
This research was supported by the DFG (SFB/TR 16), the program ``Kurzstipendien f\"ur DoktorandInnen'' of the DAAD, the Bonn-Cologne Graduate School of Physics and Astronomy, the US Department of Energy (Office of Nuclear Physics), and CONICET.}


%

}  


\end{document}